\documentclass[12pt]{article}

\usepackage{sbc-template}
\usepackage{bibentry}
\usepackage{float}

\usepackage{graphicx,url}

\usepackage[english]{babel}   
\usepackage[utf8]{inputenc}
\nobibliography*

\title{Applying Software Craftsmanship Practices to a Scrum Project: an Experience Report}

\author{Percival Lucena, Leonardo P. Tizzei }

\address{
  IBM Research\\
  São Paulo  - Brazil 
  \email{\{plucena,ltizzei\}br.ibm.com}
}

\begin{document} 

\maketitle

\begin{abstract}
  The Software Craftsmanship manifesto has defined values and principles that software development teams should follow to deliver quality software that fulfills functional and non-functional requirements without dealing with high amounts of technical debt. Software craftsmanship approach to software development prioritizes technical practices  in order to provide a clean code base.  This work analyzes a set of practices that can be applied to a Scrum project that aims to incorporate Software Craftsmanship values. The process implementation described  may be a useful contribution for  software development teams who  also intend to implement Software Craftsmanship on their projects. \\
 \\
\end{abstract}

\section{Introduction}
The 2015 CHAOS report \cite{chaos} published by Standish Group has analyzed more than 50,000 software development projects from different sizes and complexities.  Although the projects that have adopted Agile software development methodology had a higher success rate than those that had adopted Waterfall approach, more than 40\% of Agile projects had problems related to incomplete scope, low quality, or have exceeded the estimate delivery time. In the last few years, several software development companies have adopted Agile Software development methodologies, most of them adopting the Scrum framework \cite{stateofscrum}. According to  the study published by \cite{brazil}, those companies face similar problems as described in the CHAOS report. The high number of unsuccessful Scrum projects suggests that the successful framework use requires specific conditions to be met.

In order to maximize the number of successful Agile projects, several metrics that measure some quality aspects of the code have been proposed. Technical debt is one of these metrics and it is often used in current Agile projects. It was introduced by \cite{td} to define incomplete software artifacts that barely satisfy the functional requirements. Under these circumstances, adding  features to the software, or fixing defects, requires interest by  writing more code to complete features that should already be available in prior released versions.  Technical debt is also related to software that does not meet non-functional requirements such as performance, security, and usability.

According to the Scrum Guide \cite{scrumguide}, the Product Owner  is responsible for maximizing the value of the product and the work of the development team.  The Scrum Guide itself does not define how the Product Owner should prioritize the backlog stories. In many tight schedule projects, the Product Owner may decide to maximize the number of delivered stories by reducing or even eliminating the quality assurance related tasks  and other technical activities required to implement non-functional requirements, increasing the overall technical debt. Martin Fowler has identified this  problem and described it as Flaccid Scrum~\cite{flacid}. The~\cite{stateofscrum} report has identified this issue as one of the main causes of failure of Scrum projects .  

Software Craftsmanship presents an alternative craft model that places people at the center of the software development process~\cite{mcbreen}. Although Software Craftsmanship presents useful values to tackle some of the Scrum limitations, these values are abstract, and few practical guidelines are available in the literature. Due to this lack of guidelines to adopt Software Craftsmanship values, developers often adopt \textit{ad-hoc} approaches. This paper presents an experience report that describes a set of guidelines, which include practices and tools, to adopt Software Craftsmanship values on a Scrum project. These guidelines have been applied to a real world project. Besides investigating technical practices that can help reduce technical debt, it also analyzes the impact Software Craftsmanship can bring on the organization including the role and responsibility of the team, the customers, the Product Owner and  the ethical questions and trade offs involved in delivering software with quality.  

The rest of this paper is structured as follows:  Section~\ref{sec:craft} presents basic concepts of Software Craftsmanship and Section~\ref{sec:approach} describe practices and tools that were used to adopt Software Craftsmanship values on a Scrum project. Section~\ref{sec:discussao} discusses benefits and limitations of practices and tools described in previous sections and Section~\ref{sec:conclusao} concludes and suggests future work.

\section{Software Craftsmanship Overview}
\label{sec:craft}

This section introduces the Software Craftsmanship software development approach  followed by a brief discussion of related work. 

\subsection{Background}
\label{sec:background}

\cite{mcconnell} has discussed if software development should be considered art, craft, engineering or science. Although the craftsmanship metaphor is disputable, the related  movement brought to light important discussions about the importance of adopting good technical practices as a part of the software development methodology \cite{bria2008}.  The movement has started in December 2008, when  the Software Craftsmanship Summit was held in Chicago,  Illinois  establishing a set of principles for Software Craftsmanship.  In March, 2009, after an online group conversation, Doug Bradbury wrote a summary of the general conclusions in the form of a Manifesto for Software Craftsmanship \cite{manifesto}.

\begin{table}[!htb]
\caption{Software Craftsmanship Manifesto compared to Agile Manifesto}
\label{tab:comparacao}
\begin{center}
\scriptsize	{
    \begin{tabular}{| l | l |}
    \hline
    \textbf{Software Craftsmanship} & \textbf{Agile} \\ \hline
    Well-crafted software & Working software \\ \hline
    Steadily adding value & Responding to change \\ \hline
    Community of professionals & Individuals and interactions \\ \hline
    Productive partnerships & Customer collaboration  \\ \hline
    \end{tabular}
}    
\end{center}
\end{table}

Table~\ref{tab:comparacao} presents a comparison between Agile Manifesto values the Software Craftsmanship manifesto. The first value described on the Craftsmanship manifesto offered an alternative approach for the lack of quality in Agile projects as described by \cite{martin2008} on Agile conference keynote. The manifesto also aims at extending the original Agile manifesto values by proposing new ways to deliver software, to organize teams, and to deal with customer demands.

\subsection{Related Work}
\label{sec:correlatos}

The seminal book about Software Craftsmanship \cite{mcbreen}  describes concepts that help differentiate Software Craftsmanship from traditional Software Engineering.  Although this work defines several quality assurance concepts that could be added to traditional Agile Software development, it does not provide details on how the software development teams could implement such concepts.  \cite{winter} offers a more concrete view on how to implement Software Craftsmanship on a Agile Project based on clean code concepts \cite{cleancode} \cite{2011clean}  and Extreme Programming techniques \cite{xp}.

\cite{oliveira} have applied a framework to identify and measure technical debt on Scrum projects, but the authors do not provide guidelines on how to improve the Scrum process to minimize technical debt.  \cite{brown} have analyzed technical debt causes to Scrum projects and suggest using Extreme Programming and Software Craftsmanship technical practices to the project. \cite{mushtaq} also propose integrating Extreme Programming techniques to the Scrum project in order to reduce technical debt, which is only one of the several goals of the craftsmanship manifesto. This work will discuss a broader set of aspects including software delivery, team organization and customer relationship.



\section{Software Craftsmanship Applied to Scrum}
\label{sec:approach}

This section describes how we applied software craftsmanship to a Scrum project. An overview of this Scrum project is presented in Section~\ref{sec:app}. Sections~\ref{sec:qualidade}-\ref{sec:parceria} describe how existing practices and tools have been used to adopt Software Craftsmanship. The goal is to report a set of guidelines for developers willing to implement quality code according to Software Craftsmanship values. These guidelines were developed based on the experience of developers and specificities of the aforementioned Scrum project.

\subsection{Target Application}
\label{sec:app}

An e-commerce application was developed for a telecommunications company that handles customized mobile devices through its sales channel. The application was designed to be available on web and mobile devices through its own website, but also embedded as a multi-tenant application inside third party e-commerce websites. The developer team was composed of ten developers, an architect, a Scrum master, and a Product Owner. The team members were distributed in United States and in Brazil and they developed the application in 12 2-week sprints. The application back-end was developed in Java and Ruby and the front-end was developed using AngularJS. Software developers had good knowledge of technologies involved and  the company has traditionally adopted software engineering best practices.

\subsection{Well-crafted software vs Working Software}
\label{sec:qualidade}

The Agile Manifesto \cite{agile} values working software over comprehensive documentation. Many agile teams believe architecture activities are part of the system documentation and therefore are not essential to the project.  Nevertheless, system architecture is a key activity for creating mechanisms that enable a software to meet all the non-functional requirements.  Weak architectures often result in systems with high technical debt. 

In order to incorporate good architecture practices to our Scrum project our team has adopted  Scott Ambler's Agile Model Driven Development (AMDD) \cite{ambler} which proposes to include modeling tasks to Scrum stories.  According to AMDD, models should be created before coding.   An initial domain based model was also created at Sprint 0 together with the definition of the basic architecture mechanisms.  We have combined the practice with Model Driven Architecture (MDA) tools capable of generating basic code from  models and  keep them in sync with the source code base. 

Source code quality is also an essential attribute of well-crafted software. Martin's ``clean code principles'' \cite{cleancode} is a practical guide to implement Software Craftsmanship code quality practices.  Clean code principles define coding standards, rules for unit tests creation, and help to define the system architecture.

\begin{center}
\begin{figure}[H]
\centering
\includegraphics[width=\textwidth, height=6cm]{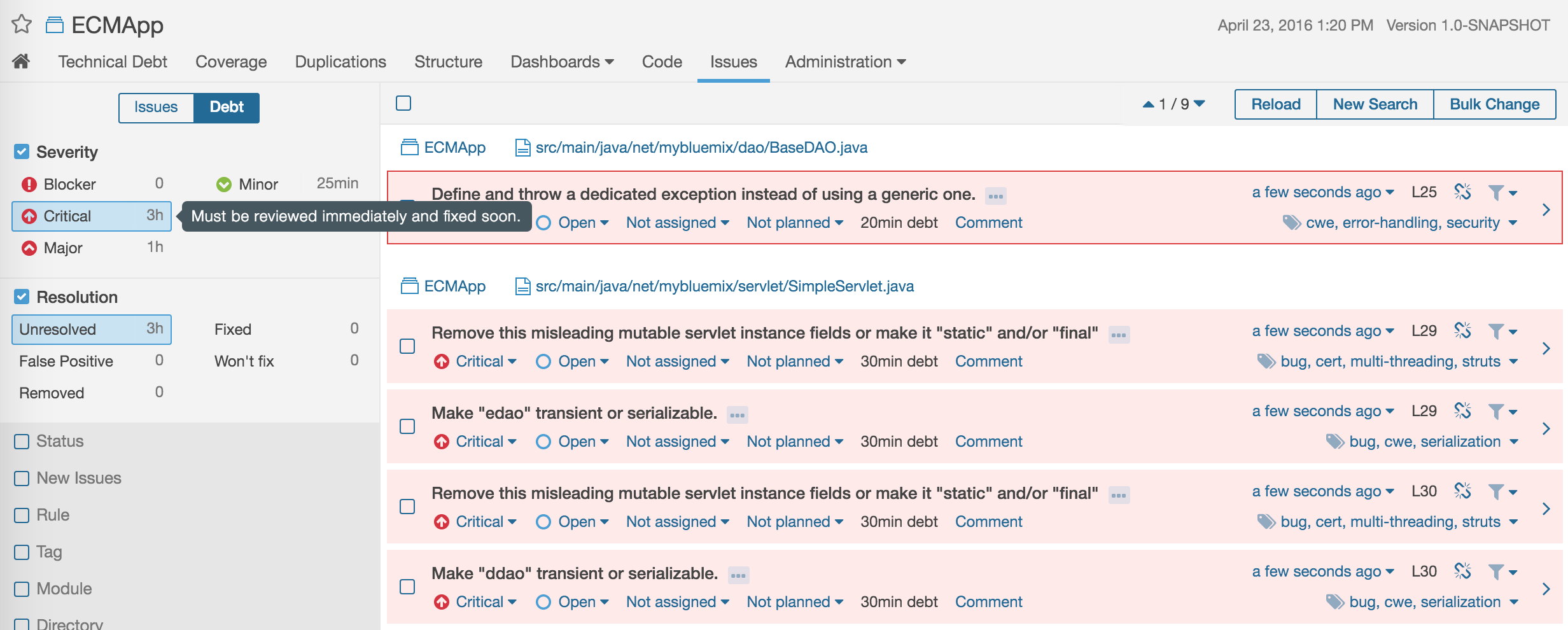}
\caption{Static Code Analysis using SonarQube }
\label{fig:sonar}
\end{figure}
\end{center}

A small subset  of the ``clean code principles'' was implemented by our team using the SonarQube static analysis tool.  This tool is based on the Squale method \cite{squale}  which defines practices as an intermediate level between quality metrics and criteria. A Squale practice abstracts the extracted code information combining and weighting different user defined metrics. Squale practices cover documentation, programming conventions, and test coverage.  The SonarQube tool were run as part of the code review process.  Figure~\ref{fig:sonar} shows standard code analysis reports generated by this tool.  The software development team has agreed to follow SonarQube Standards.  No source code pull request was accepted  in case it presented any blocker severity issue. 

Despite teams effort to eliminate technical debt, according to \cite{oliveira} a small amount of it  is inevitable in project. Since there is no simple way of eliminating technical debt completely, our project team has decided to measure and manage it.  During the project Sprints,  the gathered technical debt was measured in the project by logging hours spent on technical debt tasks separated from the hours spent on backlog tasks. Each type of task was kept on a separate backlog to help better manage it and visualize the tasks.  Although many times, the technical debt tasks could not be paid immediately, the team made some effort to slow down the technical debt growth rate.  Figure 2 shows total technical debt accumulated measured by SonarQube tool and the total numbers of hours worked on backlog stories. The gathered data suggests that team velocity slowed down as the technical debt hours grew because the team spent more hours on  technical debt tasks and less time on backlog stories tasks.

\begin{figure}[H]
\centering
\includegraphics[scale=0.3]{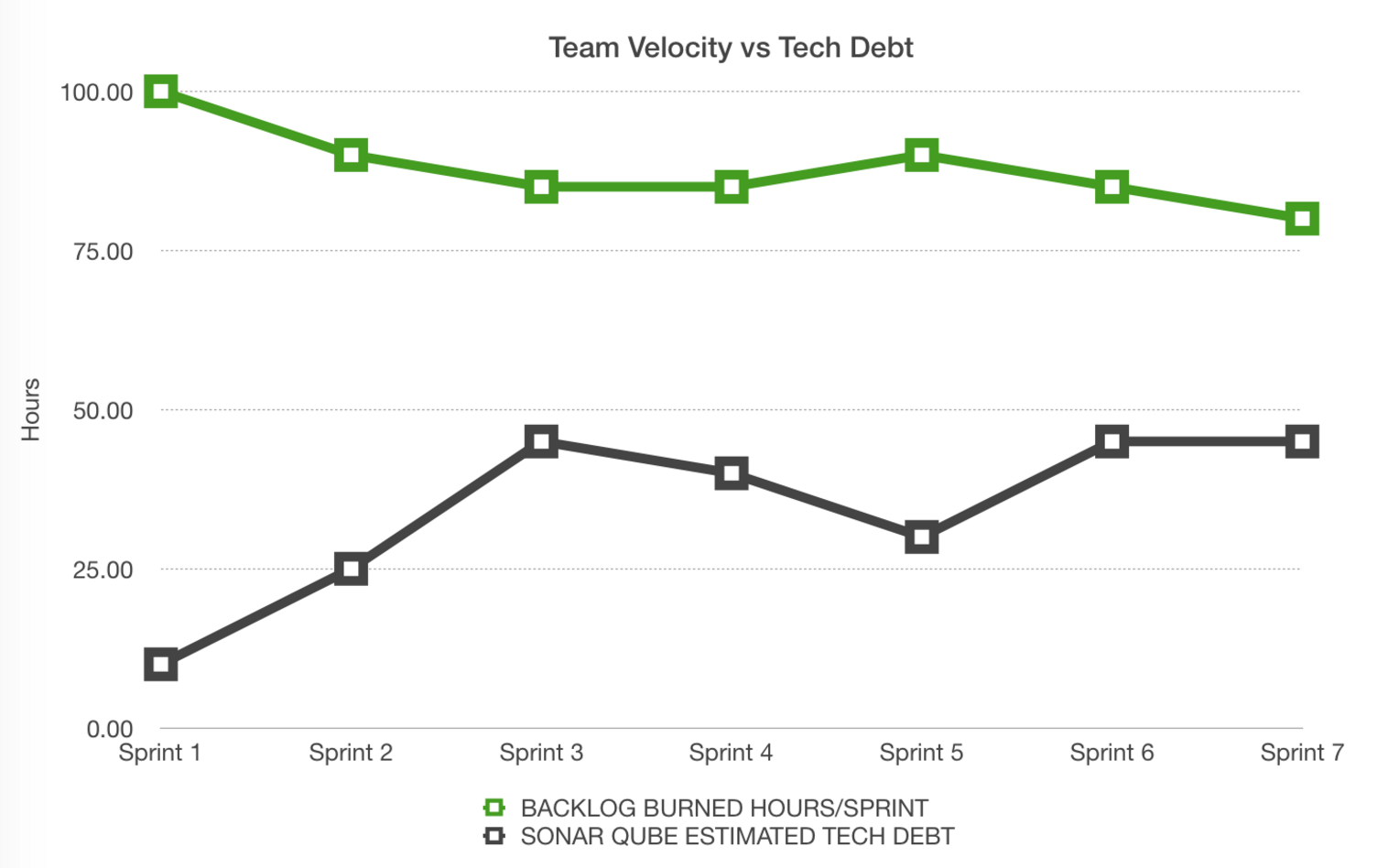}
\caption{Team Velocity measured in backlog worked hours vs SonarQube Total Technical Debt Hours}
\label{fig:cont-delivery}
\end{figure}

\begin{figure}[H]
\centering
\includegraphics[width=\textwidth, , height=4cm]{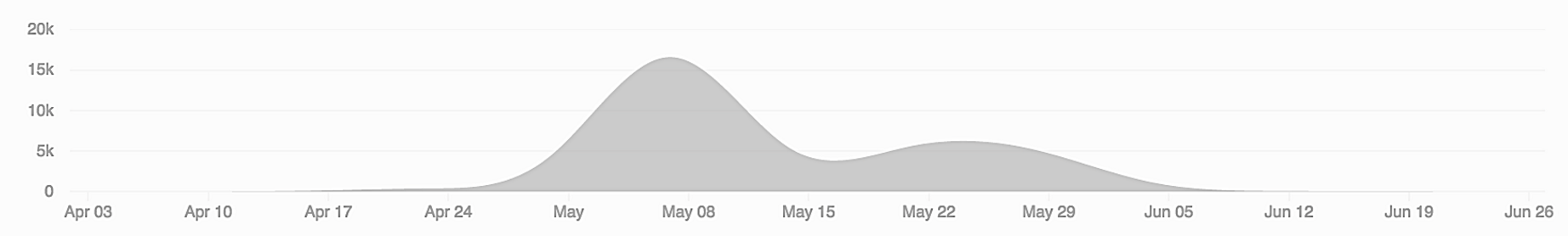}
\caption{Number of code lines deleted over time}
\label{fig:cont-delivery}
\end{figure}

Another metric used to estimate technical debt gathered was the number of code lines deleted from the git source code repository. Although some re-factoring  was often needed we have inferred that big changes were usually derived from poor planing or loose understanding of the requirements. Figure 3 shows source code deletions accumulated during a certain period of time.  Despite the fact this technical debt metric is less accurate than the one  provided by the static analysis source code tool it can be easily extracted from git repositories and can be used as an indicator to track the effort spent paying technical debt.

In order to improve source code quality and thus reduce technical debt our team has adopted a few Extreme Programming technical practices ~\cite{xp} to the Scrum based software development process. The co-located  team was organized in a way that team members could do Pair Programming by rotating positions in the room. We also have implemented a mandatory code review process using git pull requests. Once a pull request is sent to the version control system, someone who is available on team reviews the set of changes, discuss the modifications, and even push follow-up commits if necessary. During our Sprint retrospective meetings we have learned that the code review process has helped the team to improve the application architecture and fix potential defects that were not previously covered neither by our unit tests nor by SonarQube tool.

\subsection{Steadily adding value vs Responding to change}
\label{sec:valor}

Scrum's short Sprints iterations allows Product Owners to re-prioritize backlog stories, allowing the team to respond to business changes. Sprint reviews are good opportunities for the team to present the increment developed on the Sprint. Unfortunately, the software presented on Sprint reviews is sometimes executed  in hardware environments different from the customer real enterprise environment and  it includes mocks to external system interfaces. When the software is finally deployed in the real customer environment one often finds integration and performance issues which could be prevented earlier. Late software deployments also limit useful feedback for the development team and delays customer return of the investment. 

Extreme Programming has addressed this problem though the Continuous Integration technical practice.  This practice focused on  integrating and test source code changes as often as possible \cite{xp}. The Devops movement \cite{devops} has evolved this concept into  the  Continuous Delivery process that allows the software to be delivered automatically to its hosting environments. Devops adoption requires a paradigm change. The Devops movement  has worked on a set organizational  culture that removed the traditional Operation processes that manually controlled the infrastructure responsible for running IT projects.  

Due to organizational or technical constraints a large percentage of Agile teams still do not implement a continuous delivery process \cite{stateofscrum}. Continuous deployments offer constant feedback, helping the team to focus on issues that can not be foreseen by the Product Owner and that can only be discovered in real use cases.  The Devops practice is usually implemented with pipelines which consist of an automated process that executes tasks such as building the source code,  packaging the dependent libraries, run unit, integration and performance tests and deploy the resulting  software to a specific environment.  A Devops pipeline is usually composed of several deployment environments used for different proposes.

Our project team has created separate environments for development, testing, pre-production, and production. The process was automated using Jenkins Multi-Tenant as Continuous Integration tool. Git pull requests trigger our pipeline which runs a maven script to execute a Junit test-suite. In case the build breaks, the team receives a message so they can fix the broken issues. In case no test fails, a new executable package is created and  then deployed to the first development stage of the Devops pipeline.  Then, Selenium integration tests and JMeter performance tests are run against the development environment. In case all the tests succeed, the pipeline allows to move the new artifact to the remaining environments.

All the team members receive quick feedback on the results of the changes. The automated pipeline helped the team to improve the confidence level for deployments which most of times took only a few minutes to complete. The agility provided by the Devops process allowed the Product Owner to verify and use the software latest releases providing the team useful feedback about the ongoing Sprint. Production deployments occurred flawlessly providing value to the end user as soon as possible.

\subsection{Community of professionals vs Individuals and interactions}
\label{sec:comunidade}

The Agile manifesto \cite{agile} values Individuals and Iterations over processes and tools. Based on such concept, the development team could choose what practices are the best to deliver working software without wasting time with processes that do not aggregate value to the customer. Unfortunately, some Agile lightweight frameworks, such as Scrum, do not define what technical practices should be adopted \cite{agile}.  In some Scrum  projects the Scrum Master and the Product Owner may also interfere in the project organization, preventing the software development team to self organize and decide which software development process should be followed. In such case,  a professional attitude is required from the software development team who should commit to its code of ethics, being  able to deliver quality software, no matter how the enterprise is organized and what internal and external demands are in place.


\cite{sandro} suggests that Software Craftsman practitioners should  embrace a code of ethics to guarantee professionalism in software development activities.  The code of ethics should protect the development team with principles that would be strictly followed under any circumstances.  By following a code of ethics, self organizing teams should be able to define their development process with lesser external interference. A cultural change in relationship with customers is also necessary. Software developer teams should not be intimidated to change their set of useful technical  practices despite of any external pressures on their work.



Our project team has adopted the  ACM  Code of Ethics \cite{ethics} as the base for our nonnegotiable principles and work rules. This code of ethics shares several points in common with Software Craftsmanship professionalism values including the commitment to achieve the highest quality of work, acquire, and maintain professional competence and to be honest and trustworthy.



\subsection{Productive partnerships vs Customer collaboration}
\label{sec:parceria}

The Agile Manifesto \cite{agile} values customer collaboration over contract negotiation. In traditional software development contracts we have triple constraints: cost, time, and schedule. The Scrum framework provides a prioritized backlog that changes over the Sprints executions. It is not feasible to change the contracts, every time the Product backlog changes. Traditional software development contracts should have a big slack in order to accommodate those changes.  

Time and material contracts are arrangements under which the software development contractors are paid  on the basis of the worked hours agreed upon fixed add-on to cover the contractor's overheads and profit. Time material contracts are used on other industries who require open scope activities and fit well to backlog changes required on Agile based development. By establishing a flexible work model the software development team can work together with customers implementing valuable stories as needed without major concerns about exceeding fixed budgets and schedules.

On top of collaborating with the customer, the software development team should establish a productive partnership with the customer which requires engagements from the mutual parties. For the sake of successfully applying Software Craftsmanship values to a Scrum project, one needs to rethink the relationship among the software development team, the management and the customers.  Extreme Programming offers the Planning Game technical practice to help the software development team to prioritize the backlog together with the Product Owner. The software development team estimates candidate stories for the next Sprint so the  Product Owner can choose among the most valuable set of stories that can be added to the Sprint. Estimates are created  as late as possible so they are based on the best possible information.  The planning game has helped our software development team to have a closer partnership with our customer by providing the most valuable stories to the Sprints.

Towards the goal of creating a valuable product for the customer the software development team should also be committed  to understand the customer business to provide the best solution for the problem. Backlog stories written by a single Product owner might not provide all the necessary information about the business scenarios. Domain driven design \cite{ddd} proposes that software developers and the customer Domain Experts  should collaborate to create an accurate description and model of the  domain problems. This cultural change involves a true partnership between the customer organization and the software development team so they can work together and share responsibilities about the project. Domain Driven Design also require the software developers to speak a  ubiquitous language so the software model represents the same  business concept on their bounded context. Domain Driven Design was implemented by our team as part of the Sprint Planning and also as part of  Agile Modeling tasks added to the stories implementation. We found that those extra activities helped to reduce technical debt because stories details would be defined before hand reducing the re-factoring tasks required by incomplete or inaccurate requirements and models.

\section{Discussion}
\label{sec:discussao}

\cite{jacobson} consider Scrum  an incomplete software development methodology as it does not provide technical practices required by the software development team to create a quality product.  As a result Scrum teams should define their own set of technical practices based on their needs. As discussed on previous sections, our team has identified a set of  practices congruent to  Software Craftsmanship values that complement the Scrum framework.  Table~\ref{tab:tecnicas}  summarizes the technical practices adopted by our team and the life cycle phase in which they were used. Different development teams may find other technical practices more suitable for their working context. 

\begin{table}[!htb]
\caption{Software Craftsmanship Practices added to the Scrum Process}
\label{tab:tecnicas}
\begin{center}
\scriptsize	
    \begin{tabular}{| l | p{7.2cm} | l |}
    \hline
    \textbf{\#} & \textbf{Technical Practice} & \textbf{Life Cycle Phase} \\ \hline
    1 & Code of Ethics & All phases \\ \hline
    2 & Planning Game & Planning \\ \hline
    3 & Agile Modeling & Development \\ \hline
    4 & Domain Driven Design  & Development \\ \hline
    5 & Coding Standards & Development \\ \hline
    6 & Static Code Analysis & Development\\ \hline
    7 & Code Review & Development  \\ \hline
    8 & Pair Programming & Development \\ \hline
    9 & Automated Unit Testing & Testing \\ \hline
    10 & Automated Integration Testing & Testing \\ \hline
    11 & Automated Performance Testing & Testing \\ \hline
    12 & Automated Build and Deploy & Deploy \\ \hline
    \end{tabular}
\end{center}
\end{table}

We have evaluated  our practices regarding the technical debt gathered through the Sprint.  We have kept technical debt tasks separated from the Sprint backlog stories in order to understand the impact of those tasks in the Sprint execution. Static code analysis and code re-factoring measures provided useful indicators for the accumulated technical debt.  This approach can be followed by other project teams who need to track the impact of adopting a specific set of technical practices in their projects. 


\section{Conclusions}
\label{sec:conclusao}

Software craftsmanship goal is to improve existing agile manifesto principles raising the bar of  quality level delivered to software development projects. This paper has discussed a set of technical practices that can be added to the Scrum project in order to implement all the software craftsmanship manifesto values.  Our approach has extended the work proposed by \cite{mushtaq} that incorporated Extreme Programming practices to Scrum. Our approach has added more practices  from different sources including  Agile modeling,  Devops,  and from Software Craftsmanship.  The combination of all those different practices have helped our team to define requirements more accurately through agile modeling, improve the source code quality through code standards, code reviews and static code analysis, and  do extensive automated testing and deployment.

Although the team velocity at the first Sprints was smaller than similar projects who had adopted fewer technical practices, we have noticed that our team had a more stable velocity through the project. Overall, the team and was able to deliver a more valuable project to the customer with a higher quality than previous projects.


\begin{small}  
\bibliographystyle{sbc}

\end{small}


\end{document}